\documentclass[useAMS,usenatbib]{mn2e}

\usepackage{graphicx}

	\usepackage{hyperref}
	\hypersetup{colorlinks=true,citecolor=blue,linkcolor=blue,filecolor=black,runcolor=black}

\usepackage{times}

\usepackage{booktabs}

\usepackage{aas_macros}

\title[Radio source populations in the E-CDFS]{The sub-mJy radio sky in the Extended Chandra Deep Field South: source population}

\author[Bonzini et al.]{M. Bonzini\thanks{E-mail:mbonzini@eso.org}$^1$, P. Padovani$^1$, V. Mainieri$^1$, K.I. Kellermann$^2$, N. Miller$^3$, P. Rosati$^1$,
\newauthor P. Tozzi$^4$, S. Vattakunnel$^5$\\
$^1$ European Southern Observatory, Karl-Schwarzschild-Strasse 2, D--85748 Garching, Germany\\
$^2$ National Radio Astronomy Observatory, 520 Edgemont Road, Charlottesville, VA 22903-2475, USA\\
$^3$ Department of Mathematics and Physical Sciences, Stevenson University, 1525 Greenspring Valley Road, Stevenson, MD  21153-0641, USA\\
$^4$ INAF - Osservatorio Astrofisico di Arcetri, Largo E. Fermi, I-50125, Firenze, Italy\\
$^5$ INAF - Osservatorio Astronomico di Trieste, via G.B. Tiepolo 11, I-34131, Trieste, Italy}

\begin{document}

\date{May2013}

\pagerange{\pageref{firstpage}--\pageref{lastpage}}  \pubyear{2013}

\maketitle

\label{firstpage}
\begin{abstract}
The sub-mJy radio population is a mixture of active systems, that is star forming galaxies (SFGs) and active galactic nuclei (AGNs). We study a sample of 883 radio sources detected at 1.4 GHz in a deep Very Large Array survey of the Extended Chandra Deep Field South (E-CDFS) that reaches a best rms sensitivity of 6$\mu$Jy.
We have used a simple scheme to disentangle SFGs, radio-quiet (RQ), and radio-loud (RL) AGNs based on the combination of radio data with \textit{Chandra} X-ray data and mid-infrared observations from \textit{Spitzer}.
We find that at flux densities between about 30 and 100$\mu$Jy the radio population is dominated by SFGs ($\sim 60\%$) and that RQ AGNs become increasingly important over RL ones below 100 $~\mu$Jy. 
We also compare the host galaxy properties of the three classes in terms of morphology, optical colours and stellar masses.
Our results show that both SFG and RQ AGN host galaxies have blue colours and late type morphology while RL AGNs tend to be hosted in massive red galaxies with early type morphology.
This supports the hypothesis that radio emission in SFGs and RQ AGNs mainly comes from the same physical process: star formation in the host galaxy. 
\end{abstract}


\section{Introduction}
\label{Intro}

The two main processes that contribute to the extragalactic continuum radio emission at 1.4 GHz are the non-thermal emission associated with relativistic electrons powered by active galactic nuclei (AGNs) and the synchrotron emission from relativistic electrons in supernova remnants. The latter is therefore a tracer of star formation activity in galaxies. 
While the bright radio sky is dominated by the emission driven by ``radio-loud'' AGNs, at fainter flux densities ($<$ 1 mJy) the contribution from star forming galaxies (SFGs) become increasingly important \citep[e.g.,][]{prandoni01,smolcic08,seymour08,padovani09}. 
Moreover, recent work has revealed the presence of a third population of sources in the sub-mJy radio sky, the ``radio-quiet'' AGNs \citep[e.g.,][]{padovani09}. These sources show the presence of AGN activity in one or more bands of the electromagnetic spectrum (e.g., optical, mid-infrared, X-ray) but the origin of their radio emission has been a matter of debate.
It has been proposed that they represent scaled versions of RL AGNs with mini radio jets \citep[e.g.,][]{giroletti09} or that their radio emission comes from star formation in the host galaxy \citep[e.g.,][]{padovani11b,kimball11}.
Disentangling the two emission mechanisms is important to investigate the circumstances under which they originate, for example the host galaxy properties, 
and possibly to study the connection between AGN and star formation activity.
Most of the radio sources in surveys as deep as the one discussed in this paper are barely resolved or unresolved. So radio observations alone generally can not be used to distinguish between AGN and star formation driven sources; hence, the need for a multi-wavelength approach. 
For example, \citet{padovani09} showed that the ``standard'' definitions of radio-loudness, which are based on radio-to-optical flux density ratios, R, and radio powers, are clearly insufficient to identify radio-quiet AGN in faint radio samples. 
This is because these also include star-forming and radio-galaxies;  both of these classes are or can be, respectively, characterized by low R and low radio powers as well. \citet{padovani11b} showed instead that a proper classification of faint radio sources  requires a combination of radio, IR, and X-ray data. 
Here, we propose a new simple classification scheme, which is an upgrade of that used by \citet{padovani11b}.
The paper is organized as follows. We present our sample and the data used for the classification of the radio sources in Section \ref{sec_sample_and_data}. Our classification scheme is presented in Section \ref{sec-sfg-vs-agn}. In Section \ref{sec_number_counts} we describe the relative contribution of the different source types to the radio population. The host galaxy properties are analysed in Section \ref{sec_host_gal}. 
In Section \ref{sec_discussion} and \ref{sec_summary} we discuss and summarize our results. 
Our definitions of AGN, RQ, and RL sources follow those of \citet{padovani11b}. 
In this paper we use magnitudes in the AB system, if not otherwise stated, and we assume a cosmology with $H_{0}=70$ km s$^{-1}$ Mpc$^{-1}$, $\Omega_{M}=0.3$ and $\Omega_{\Lambda}=0.7$.

\section{Sample and data}
\label{sec_sample_and_data}

\subsection{Radio catalog, optical-IR counterparts and redshifts}
\label{sec_radio_sample}
We consider a sample of 883 radio sources detected at 1.4 GHz in a deep VLA survey of the Extended Chandra Deep Field South (E-CDFS) that reaches a best rms sensitivity of 6 $\mu$Jy. 
The average 5$\sigma$ flux density limit is 37 $\mu$Jy and the spatial resolution is $2.8\arcsec \times 1.6\arcsec$.
A description of the survey strategy and the data reduction details are given in \citet{miller13}. 
Using the wealth of optical and infrared (IR) data available in the E-CDFS we were able to identify the optical/IR counterpart for 839 (94\%) radio sources using a likelihood ratio technique \citep{bonzini12}. 
Combining data from the literature and newly acquired spectra, we assigned a reliable spectroscopic redshift to 274 sources. Including photometric redshift the number of sources with measured redshift increases to 678 and their average redshift is $z\sim1.1$. The accuracy of the photometric redshift is around 6\% \citep{bonzini12}.
The majority of the objects without a redshift are not detected at optical wavelengths. 
Their counterparts can be detected only at longer wavelengths, in the near or mid-infrared. Therefore, spectroscopic observations are challenging and photometric estimates, based on a few photometric points, are not robust.
In other cases, the lack of photometric redshift information is due to a less rich multi-wavelength coverage, especially in the outskirts of the field.  

\subsection{Mid infrared data}
Mid-infrared (MIR) wavelengths can be used to unveil the presence of an AGN (see Section \ref{sec-IRAC-colours}). Therefore, we used deep \textit{Spitzer} IRAC and MIPS data.
The IRAC data were obtained as part of the SIMPLE survey \citep{damen11}. It covers an area of about 1,600 arcmin$^2$ centred on the E-CDFS. 
The typical 5$\sigma$ flux density limits are 1.1, 1.3, 6.3, and 7.6 $\mu$Jy at 3.6, 4.5, 5.8, and 8.0 $\mu$m, respectively.
We also use MIPS 24$\mu$m data from the Far-Infrared Deep Extragalactic Legacy Survey (FIDEL) \citep{dickinson07}.
This survey covers, at 24$\mu$m, $\sim$90\% of the VLA area considered in this paper and the typical flux density limit (5$\sigma$) is 30 $\mu$Jy. 
To associate the correct MIR photometry to our radio sources we cross-correlated the position of their optical/IR counterparts with the SIMPLE and FIDEL catalogues, after correcting for the median offset in right ascension and declination between the two samples. 
The matching radii are 0.7$\arcsec$ for the SIMPLE catalogue and 1.5$\arcsec$ for the FIDEL catalogue.
A total of 91\% and 88\% of the identified radio sources (839) have a match in the SIMPLE and FIDEL catalogue, respectively.
For sources without 24$~\mu$m detection we compute an upper limit to the 24$~\mu$m flux density. Since the exposure time varies across the field the upper limit is extrapolated from the exposure map as: Flux$_{lim}(1\sigma)= -1.4+1369.1(1/t_{exp})$, where $t_{exp}$ is the exposure time in seconds, and the flux density is in $\mu$Jy.

\subsection{X-ray data}
\begin{figure}
	\centering
	\includegraphics[width=\columnwidth]{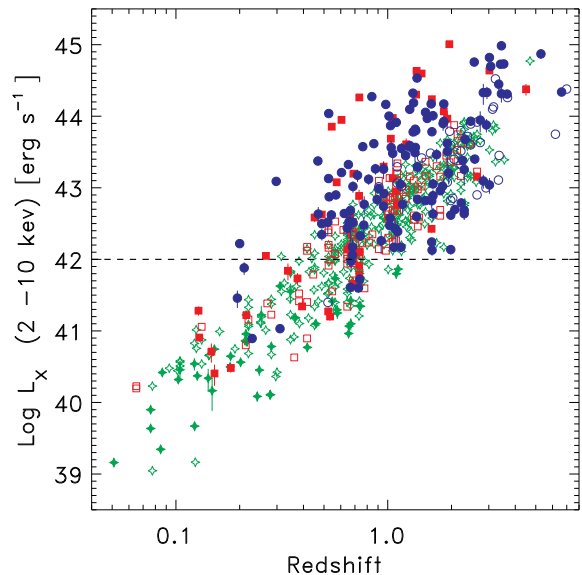}
 \caption{\small{X-ray luminosity as a function of redshift for RL AGNs (red squares), RQ AGNs (blue circles), and SFGs (green crosses) (See Section \ref{scheme}). Filled symbols represent X-ray detections, while open symbols are 3$\sigma$ upper limits. X-ray detected sources with luminosity above the horizontal line are considered to be AGNs.}}
 \label{fig_Lx-vs-z}
\end{figure}

The E-CDFS has been mapped in the X-ray band by \textit{Chandra}. 
A total of 129 radio sources have a counterpart in the 4 Ms observations of the CDFS presented in \citet{xue11} and another 99 in the main E-CDFS catalogue by \citet{lehmer05} obtained with shallower (250 ks) observations in each of four pointings.
The list of the X-ray counterparts of the radio sources is given in  \citet{bonzini12}. 
We compute the X-ray luminosity (2-10 keV) to indicate the presence of AGN activity (see Section \ref{sec-X-criteria}). 
Objects without X-ray counterpart, but with redshift (435 sources), have only an upper limit (3$\sigma$) on the X-ray luminosity obtained from aperture photometry on the X-ray images at the position of the radio source \citep{vattakunnel12}.
A total of 44 radio sources lie in the outermost part of the field and have no X-ray observations available.
The X-ray luminosity for our sample as a function of redshift is plotted in Fig. \ref{fig_Lx-vs-z}.

\section{AGN or SFG?}
\label{sec-sfg-vs-agn}
As discussed in Section \ref{Intro}, a proper classification of the faint radio sources requires a wealth of multi-wavelength data.
We have identified three diagnostics that can be used to split the population in the different classes of sources. In this section, we first describe these diagnostics and how they are used to classify the sources (Section \ref{criteria}) and our classification scheme (Section \ref{scheme}). Then, we present a series of checks that we performed to validate the method or to refine the classification of some peculiar sources (Sections \ref{sec_src_wo_z}, \ref{sec-unidentified}, and \ref{sec_checks}). 

\subsection{Classification criteria}
\label{criteria}

\subsubsection{q-values}
\label{sec-q-value}
It is well known that the far-IR and radio emission are tightly and linearly correlated in star-forming systems \citep[e.g.,][and references therein]{sargent10}. This is usually expressed through the so-called $q$ parameter, that is the logarithm of the ratio of far-IR to radio flux density, as defined by \citet{hel85}. 
Ideally, one would like to derive a bolometric $q$,  but typically insufficient data are available at longer wavelengths to do this reliably. 
For example, \citet{padovani11b} were able to derive only upper limits on $q$ for 50\% of the sources. 
In this paper, we therefore prefer to use $q_{24obs}$, which is defined as: 
\begin{equation}
q_{24obs}=log_{10}(S_{24\mu m}/S_{\rm r}),
\end{equation}
where $S_{24\mu m}$ is the observed 24$~\mu$m flux density, and $S_{\rm r}$ is the observed 1.4 GHz flux density.
The use of the observed flux densities rather than the rest frame ones minimises the uncertainties due to the modelling.
In case of resolved sources, we use integrated radio flux density \citep[see][for details]{miller13}. 
The distribution of the $q_{24obs}$ values as a function of redshift is plotted in Fig. \ref{fig_q24_vs_z_pop}.
For a given intrinsic spectral energy distribution (SED), this ratio has a redshift dependence.
We assume that the IR and radio properties of high redshift SFGs are similar to the local ones \citep[e.g.][]{sargent10}. 
Therefore, to define a locus of SFG, we compute $q_{24obs}$ as a function of $z$ using the SED of M82\footnote{From the SWIRE library, \citet{polletta07}} as representative of the SFGs class from $z=0$ to the maximum redshift of our sources. 
For the radio spectra we assume a spectral index $\alpha_{\rm r}$\footnote{We define the spectral index as: $S_{\nu}\propto \nu^{-\alpha} $} of 0.7, as expected for a typical SFG \citep[e.g.,][]{ibar10}.
The M82 template is normalized to the local average value of $q_{24obs}$ as obtained in \citet{sargent10} ($\langle q_{24obs} \rangle =1.31^{+0.10}_{-0.05}$ for sources with $0.08<z<0.23$).
The average spread for local sources is 0.35 dex \citep{sargent10}.
We define the SFG locus as the region of $\pm2\sigma$ centred on the M82 template (see Fig. \ref{fig_q24_vs_z_pop}).
Sources below this locus display a radio excess and therefore do not follow the far-IR -- radio correlation and are classified as RL AGNs.   
Sources without 24 $\mu$m detection have only upper limits on the $q_{24obs}$ value. These sources are classified as RL AGN if their upper limit is smaller than the M82 template at the source redshift  $-1\sigma$ (rather than $-2\sigma$). 

For a small fraction (16\%) of the radio sources 70$\mu$m \textit{Spitzer} photometry from the FIDEL survey is also available \citep{dickinson07}. We can therefore check if we obtain the same classification using the longer wavelength data rather than the 24$\mu$ flux density.
Following the same procedure described above, we compute the $q_{70obs}$ value for this sub-sample and define the corresponding SFGs locus. 
We find an excellent agreement (96\% of the cases) which validates our use of the 24 $\mu$m data for most of the sources. 

\subsubsection{X-ray luminosity}
\label{sec-X-criteria} 
Sources with hard band X-ray luminosity above $10^{42}$ erg s$^{-1}$ are considered to be AGN driven \citep[e.g.][]{szokoly04}. 
A total of 162 sources in our sample have X-ray detection above this threshold.
When only upper limits are available in the X-ray, we assume that the source is a SFG if there is no indication of black hole driven activity in others bands (e.g. IRAC colours or $q_{24obs}$). 
In the central part of the field, where the 4Ms \textit{Chandra} observations are available, the upper limits on the X-ray luminosity are so faint that we miss only the most absorbed AGN. 
In the outer part of the ECDFS we only have the 250ks \textit{Chandra} observations and therefore it is possible that we are not sensitive even to moderate luminosity ($10^{42}$ erg s$^{-1}<L_{\rm x} <10^{44}$ erg s$^{-1}$) AGN.
To avoid misclassifying sources as SFGs, sources with upper limits on their flux density $2\sigma$ above the local background level and with upper limits on their X-ray luminosity minus $1\sigma$ larger than $10^{42}$ erg s$^{-1}$ are classified as AGN. A total of 23 sources satisfy this requirement.

\subsubsection{IRAC colour-space}
\label{sec-IRAC-colours}
\begin{figure}
	\centering
	\includegraphics[width=\columnwidth]{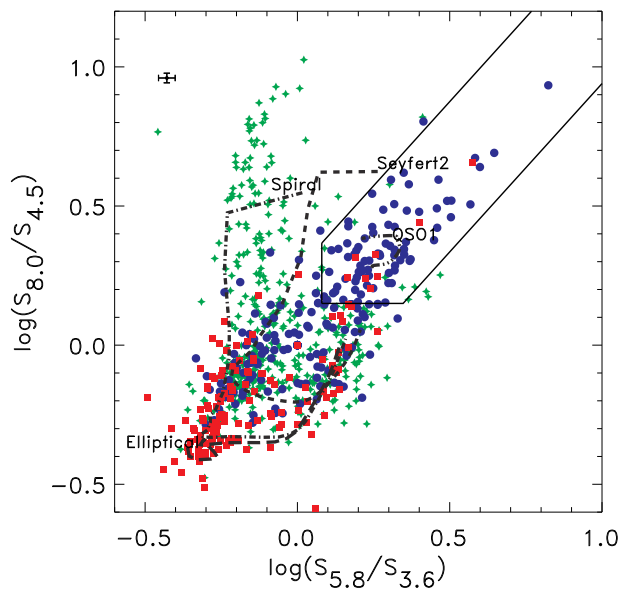}
 \caption{\small{IRAC-colours plot for RL AGNs (red squares), RQ AGNs (blue circles), and SFGs (green crosses). The black line encloses the "Donley wedge" \citep{donley12}; sources whose AGN emission dominates the MID-IR populate this region. The black cross on the top left shows the typical uncertainties.
Colour-colour tracks of a 13 Gyr old elliptical galaxy (long dashed line), a spiral galaxy (dot-dashed line), a Seyfert 2 galaxy (dashed line), and a Type 1 QSO (dot-dot-dashed line) \citep{polletta07} in the redshift range 0.1--3 are also plotted. }}
 \label{fig_IRAC_colors_pop}
\end{figure}

Finally, we use the IRAC colour-colour diagram to help select AGNs.
The AGN emission can heat up the surrounding dust that re-emits this energy in the MIR. If the AGN is sufficiently luminous compared to its host galaxy, the emission from the heated dust can produce a power-law thermal continuum across the four IRAC bands.
Sources with this spectral shape occupy a specific region in the IRAC colour-colour diagram, the so called "Lacy wedge" \citep{lacy04}.
However, the "Lacy wedge" is heavily contaminated by high redshift SFGs \citep{donley12}. 
Therefore, to select AGNs we adopt the stricter criteria described in \citet{donley12} that are designed to minimize the contamination from both low and high redshift SFG. These criteria require the flux density to monotonically increase in the four IRAC bands and the colours to be such that the source lies in the wedge plotted in Fig. \ref{fig_IRAC_colors_pop}.  
The completeness of this AGN selection method is strongly dependent on the AGN luminosity, being high 
for $L_{2-10 keV} \geq 10^{44}$ erg s$^{-1}$ but relatively low ($\la 20\%$) at lower luminosities \citep{donley12}. Seyfert 2 galaxies, are also easily missed by this diagram as show by the colour-colour track in Fig. \ref{fig_IRAC_colors_pop}.
This method is also incomplete for luminous AGN with heavy obscuration and particularly bright host galaxy \citep{donley12}.  
A total of 85 sources in our sample satisfy these criteria. 
Eight of them were already classified as RL AGN due to their $q_{24obs}$ value and about half of them (44) have also X-ray luminosity above $10^{42}$ erg s$^{-1}$. But 39 sources are classified as AGN only because of their IRAC colours.

\begin{figure}
	\centering
	\includegraphics[width=\columnwidth]{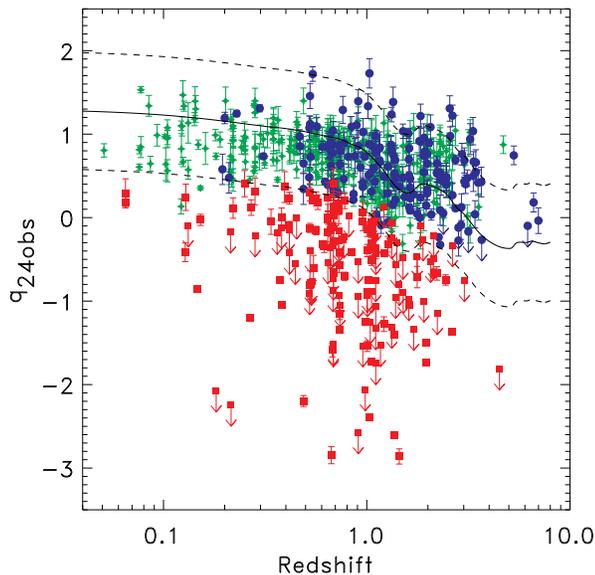}
 \caption{\small{$q_{24obs}$ values as a function of redshift for RL AGN (red squares), RQ AGN (blue circles), and SFG (green crosses). Down-pointing arrows represent 3$\sigma$ upper limits. The solid line show the evolution of $q_{24obs}$ for the M82 template as a function of redshift with $\pm 2\sigma$ dispersion (dashed lines).}}
 \label{fig_q24_vs_z_pop}
\end{figure}

\subsection{Our classification scheme}
\label{scheme}
\begin{itemize}
\item We classify as RL AGNs all sources with $q_{24obs}$ below the SFGs locus (see section \ref{sec-q-value} for details).
\item  Above this threshold, a source is classified as a RQ AGN if it shows clear evidence for an AGN in the X-ray (see Section \ref{sec-X-criteria}) or in the MIR bands (see Section \ref{sec-IRAC-colours}).
\item Otherwise a SFG classification is adopted.
\end{itemize}

\subsection{Sources without redshift}
\label{sec_src_wo_z}
In our sample, there are 161 identified radio sources without redshift. 
Only the IRAC colour criterion can be used without knowing the redshift as both the classification based on the $q_{24obs}$ value and the X-ray luminosity require this information.
However, in the following cases a reliable classification is possible even for sources without redshift:
\noindent
(i) Five sources whose $q_{24obs}$ value (detection or upper limit) is below the SFG locus defined in Section \ref{sec-q-value}, even assuming a high redshift (namely, $z=3$), are classified as RL AGNs. Indeed, given the SED shape of SFGs like M82, the threshold value between RL AGNs and RQ AGNs-SFGs decreases with redshift. 
Therefore, the assumption of high redshift for the sources without $z$ provides a conservative estimate for the number of additional RL AGNs. 
\noindent
(ii) When the IRAC flux densities follow the \citet{donley12} criteria we classify the source as an AGN. 
To distinguish between RL and RQ AGNs we use again the $q_{24obs}$ value (detection or upper limit) assuming z=3.
A total of 26 RQ AGNs without z are identified using their IRAC colours.

\noindent
(iii) Eleven sources without measured redshift are detected in the X-ray band. 
They all lie in the region with only 250ks of \textit{Chandra} observations.
Even assuming the mean redshift of the sample ($z=1$) their X-ray luminosity in the total band is greater than $10^{42}$ erg s$^{-1}$ and therefore we classify them as AGNs.
None of them are classified as RL AGN based on their upper limit on the $q_{24obs}$ value.
Most of them are in the ``Donley" wedge too, but four are ``new" RQ AGNs.
\noindent
For all the other 126 identified radio sources without measured redshift we do not have enough evidence to reveal the presence of an AGN (only upper limits on the X-ray luminosity, not in the ``Donley wedge", and not below the SFG locus) and therefore we conservatively consider them as SFGs. Note that a fraction of these could be RQ AGNs.

\subsection{Unidentified sources}
\label{sec-unidentified}
 There are 44 sources without any optical, IR or X-ray counterpart. 
For these, the only information we have is the radio flux density. All but four of them (RID 1, 78, 853, and 865) have been observed by the FIDEL survey and hence we have an upper limit on their 24 $\mu$m flux density. 
Having no information about the source redshift we assume z=3 (see discussion in Section \ref{sec_src_wo_z}) to find additional RL AGNs. 
Out of the 40 sources for which we can compute an upper limit on the $q_{24obs}$, we classify four more objects as RL AGNs.\\

Note that sources without redshift and unidentified sources are not included in the host galaxy properties analysis presented in Section \ref{sec_host_gal} since no reliable morphology, stellar mass, and colours can be derived for these sources. Therefore, the possible overestimation of the number of SFGs discussed above should not effect our results. 

\subsection{Further checks}
We have checked our classifications using additional data such as radio observations at other frequencies and optical and X-ray spectra.
These checks are summarized below, from longer to shorter wavelength.
\label{sec_checks}

\begin{itemize}
\item Inverted radio spectra sources:\\
The E-CDFS has been also observed in the ATLAS 5.5 GHz survey \citep{huynh12}. This survey has an almost uniform sensitivity of $\sim$12 $\mu$Jy rms with a resolution of 4.9 arcsec $\times$ 2.0 arcsec. In the central part of the field, deeper VLA observations at 4.8 GHz are available, down to a rms noise of 7 $\mu$Jy  \citep{kellermann08}.
We combine these sets of data to compute the radio spectral index, $\alpha_{\rm r}$, of our sources.
We use the VLA measurements in the central part of the field and the ATLAS ones outside this region. A discrepancy between the flux density measured in the VLA and ATCA surveys has been noted by \citet{huynh12}. The ATLAS flux densities are about 20 per cent greater than the VLA ones and they have been therefore corrected before computing the spectral indexes.
We measured the radio spectral index for a total of 215 sources excluding the multiple component sources as the interpretation of their spectral index is complicated by their core-jet structure.
An inverted radio spectrum ($\alpha_{\rm r}<0$) is the signature of compact core emission typical of radio AGNs \citep[e.g.,][]{kellermann69}.
Only one source (RID 640) with a reliable inverted radio spectrum has been initially classified as an SFG. It lies at the bottom of the SFGs locus shown in Fig.~\ref{fig_q24_vs_z_pop}, and has therefore been re-classified as a RL AGN.

\item VLBA sources:\\
\citet{middelberg11} have detected 21 VLA-CDFS sources with the Very Long Baseline Array (VLBA) using a resolution of $\sim 0.025^{{\prime}{\prime}}$. 
With a flux density limit of $\sim 0.5$ mJy, very long baseline interferometry (VLBI) detections above $z > 0.1$ are most likely to be due to AGN. Reassuringly, all of the 20 detected VLBA objects with $z > 0.15$ had been classified as RL AGN by our method. The single object with $z < 0.15$ ($z = 0.08$) has a VLBI detection offset from the centre of the galaxy and quite a low core radio power ($\sim 5 \times 10^{21}$ W Hz$^{-1}$) and was therefore classified  as an SFG by \citet{middelberg11}, in agreement with our classification. 

\item{Optical spectra:}\\
The presence of broad lines or high excitation emission lines in optical spectra is another indicator of AGN activity. Therefore, we inspected the spectra of the sources that we classify as SFGs, when available.
Out of the 100 spectra that we inspected, none have broad emission lines and only one (RID 618) shows a high excitation emission line (NeV).  
Therefore, we changed its classification to a RQ AGN.
Many diagnostic methods based on line ratios, such as the Baldwin, Phillips \& Terlevich (BPT) diagram \citep{baldwin81}, have been proposed in the literature \citep[e.g.,][]{kauffmann03,smolcic06,best12}. The spectral coverage of the optical spectra makes these methods feasible only for local sources ($z \la 0.3$). In our sample, only 4\% of the sources have the required redshift and a good quality spectrum available. Therefore we do not have the data for a statistically significant comparison with the spectral line ratio diagnostic methods.

\item $R$ values: \\
The ratio between the rest-frame radio-to-optical flux density ratio $R$ can also be used as an indicator of radio loudness \citep[e.g.][]{kel89}. Following \citet{padovani09} we define $R$ as the logarithm of the ratio between the rest frame flux density at 1.4 GHz and in the V band, which  means that the ``classical" dividing line between radio-loud and radio-quiet AGN is at $R \sim 1.4$.
The numerator is computed using the observed radio spectral index, where available, or assuming  $\alpha_{\rm r}=0.7$ in case of sources with no detection at 6 cm, while the K-correction for the V-band flux density is computed interpolating the observed optical photometry at the rest frame V-band wavelength.
The optical photometry for our sources is taken from \citet{cardamone10} and \citet{taylor09}. We obtained an estimate of $R$ for a total of 574 source. 
The others have no detection in the observed optical and NIR wavelength (K-band) or lack a redshift estimate. 
The $R$ values are plotted versus $q_{24obs}$ in Fig. \ref{fig_R-vs-q24}. 
At $z < 1$, SFGs and RQ AGNs are both characterized by low values of $R$, with mean of 0.5 and 0.6, respectively.
These values increase as the redshift increases. This behaviour is due to our flux density limit: at high redshift we can detect only galaxies with high star formation rate. Such sources are usually dusty and, for a given optical luminosity, have higher infrared emission.
Therefore, they can have $R$ values as high as 2 but, at the same time, follow the radio-far-infrared correlation indicating that they radio emission is due to star formation rather than to the presence of a RL AGN.
On the other hand, RL AGNs span a wide range of $R$ values at all redshift. 
Therefore we do not apply a cut in $R$ to separate RQ and RL AGN since these include radio quasars and radio galaxies. 

\begin{figure*}
	\centering
	\includegraphics[width=\columnwidth]{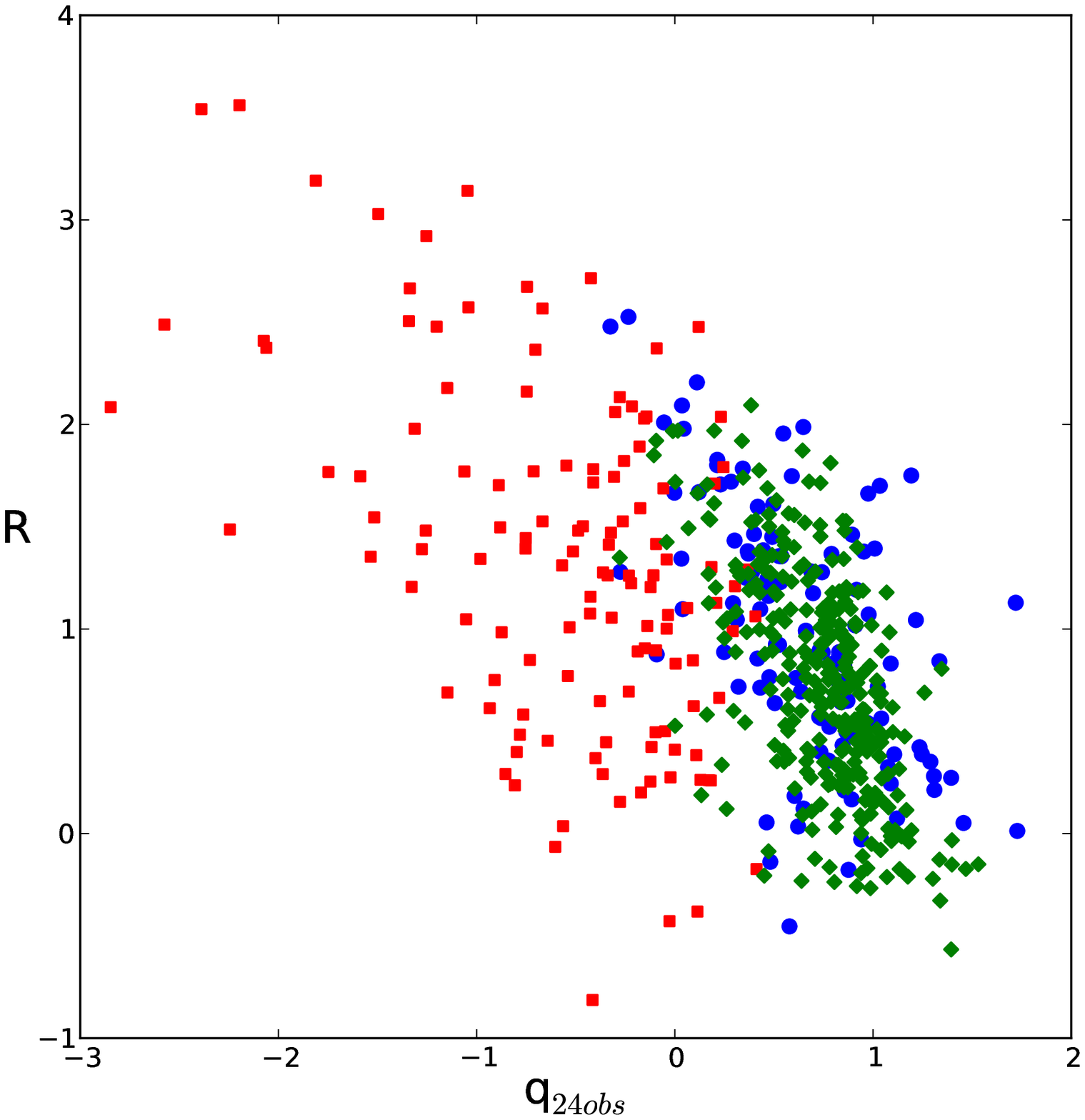}
	\includegraphics[width=\columnwidth]{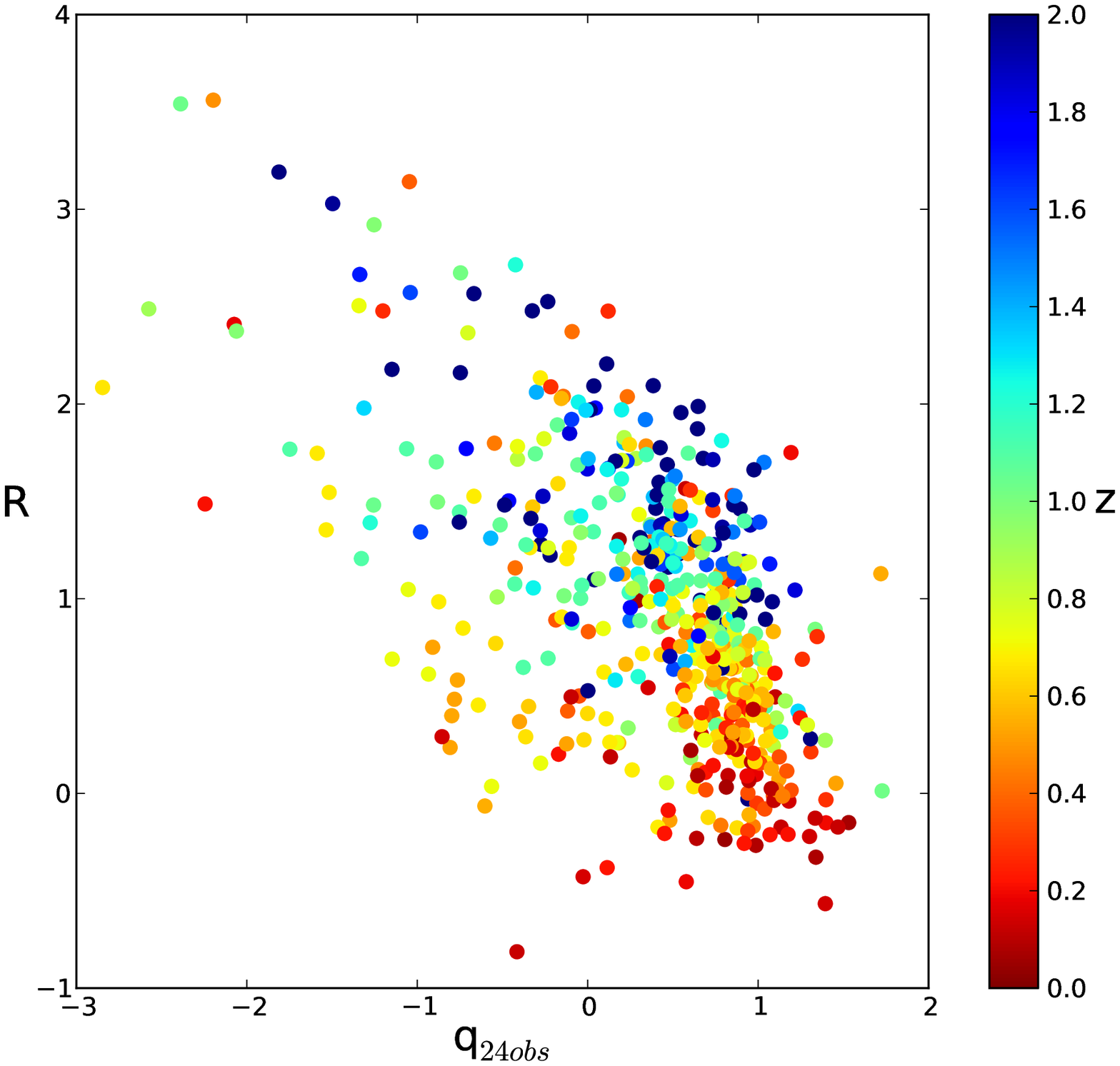}
 \caption{\small{$R$ versus $q_{24obs}$. In the left panel the different symbols correspond to the three classes of sources: RL AGN (red squares), RQ AGN (blue circles), and SFG (green diamonds). In the right panel the colour scale indicates the redshift.} 
 }
 \label{fig_R-vs-q24}
\end{figure*}

\item PAH:\\
The top left region in the IRAC colour-colour plot in Fig. \ref{fig_IRAC_colors_pop}, $Log(S_{8.0}/S_{4.5})>0.3$ and $Log(S_{5.8}/S_{3.6})<0$, is populated by sources whose SED is dominated by the Polycyclic Aromatic Hydrocarbons (PAH) features, typical of SFG. 
In particular, only low redshift SFG have IRAC colours that fall in this region since higher redshift objects tends to move to the bottom part of the plot. 
Only two (RID 568 and 438) among the $\sim 80$ sources in this region are classified as RL AGNs according to their $q_{24obs}$ value. All the others are SFGs according to our classification scheme. For these two objects it is possible that the 24 $\mu$m flux density is underestimated. Indeed, the FIDEL catalogue is obtained using aperture photometry, and for local extended sources some of the flux density might be lost. These two sources has been reclassified as SFGs (in Fig. \ref{fig_q24_vs_z_pop} they are the only two green points below the SFG locus).

\item{X-ray spectral analysis and variability:}\\
\citet{tozzi09} and \citet{vattakunnel12} performed a full X-ray spectral analysis for the X-ray detected radio sources in the E-CDFS. They considered as further diagnostics to discriminate between AGNs and SFGs the intrinsic absorption in the X-ray spectrum and the presence of the iron emission line.
Indeed, the detection of a significant intrinsic absorption reveals that the X-ray flux is dominated by nuclear emission \citep[e.g.,][]{alexander05,brightman11}. We adopt as a threshold a column density of  $10^{22}$ cm$^{-2}$.
Out of the 87 sources above these threshold, only three (ID 143, 397, and 734) were initially considered SFGs. We changed their classification to RQ AGNs. 
Another strong indicator of nuclear activity is the presence of a K-shell Fe line at 6.4 keV in the source spectra \citep[e.g.,][]{nandra94}. 
Four sources, that were initially classified as SFGs, have a clear detection of the Fe line and were therefore re-classified as RQ AGN.
Finally, the time X-ray variability is another signature of the presence of an AGN.
A total of 23 sources in our sample have X-ray variability with high confidence level ($>97\%$) \citep[][Paolillo et al. 2013, in prep.]{paolillo04}. 
They were all already classified as AGNs according to our criteria.  

\end{itemize}
To summarize, only 11 sources needed to be reclassified, supporting the validity of our classification method described in Section \ref{scheme}.
The majority of them (6/11) are RQ AGNs that were previously classified as SFGs: in such systems only the X-ray spectral analysis has revealed the clear presence of an AGN, while the other indicators were inconclusive. 
 
\subsection{Results}
According to the criteria described in the previous sections, out of the 883 radio sources, we identify 173 (19\%) RL AGNs, 208 (24\%) RQ AGNs, and 502 (57\%) SFGs.

\begin{table*}
\scriptsize
\caption{Classification of radio sources.
A full version of the Table is available in the on-line material.}
\begin{tabular}{r c c c c c c c c c c c c}
\hline
 (1) & (2) & (3) & (4) & (5) & (6) & (7) & (8) & (9) & (10) & (11) & (12) &(13) \\
 RID & RA & DEC & Class & z & Pr & Lx & $q24_{obs}$ & $log(S_{5.8}/S_{3.6})$ & $log(S_{8.0}/S_{4.5})$ & $\alpha_{\rm r}$ & B Mag & QF \\
\hline
601 &  03:32:48.18 & -27:52:56.60 & RQ AGN & 0.67 & 22.79 & 42.68 & 
0.62 & 0.53 & 0.53 & --- & 
-22.77 & 3 \\
602 &  03:32:48.30 & -27:56:26.91 & SFG & 0.35 & 22.53 & $<$41.44 & 
0.69 & 0.61 & 1.18 & 0.8 & 
-22.08 & 2 \\
603 &  03:32:48.57 & -27:49:34.36 & SFG & 1.12 & 23.69 & 41.86 & 
0.07 & 0.63 & 0.62 & 0.4 & 
-22.00 & 2 \\
604 &  03:32:49.19 & -27:40:50.49 & RL AGN & 1.22 & 25.63 & 43.61 & 
-0.42 & 2.53 & 2.76 & 0.8 & 
-23.79 & 3 \\
605 &  03:32:49.22 & -28:03:44.64 & RQ AGN & 0.64 & 22.89 & 42.44 & 
0.86 & 1.09 & 1.16 & --- & 
--- & 3 \\
606 &  03:32:49.33 & -27:58:45.19 & SFG & 2.21 & 24.84 & $<$42.96 & 
0.16 & 1.25 & 1.49 & 0.8 & 
-24.61 & 1 \\
607 &  03:32:49.39 & -27:36:36.22 & SFG & 0.64 & 22.87 & $<$42.28 & 
1.03 & 0.75 & 0.92 & --- & 
-22.31 & 1 \\
608 &  03:32:49.42 & -27:42:35.14 & RL AGN & 0.98 & 25.36 & $<$42.12 & 
-1.25 & 0.56 & 0.78 & 1.3 & 
-22.33 & 3 \\
609 &  03:32:49.92 & -27:34:45.69 & SFG & 0.25 & 22.59 & $<$41.20 & 
0.95 & 0.79 & 4.37 & --- & 
-21.16 & 3 \\
610 &  03:32:49.95 & -27:34:32.74 & SFG & 0.25 & 22.86 & $<$41.11 & 
0.81 & 0.75 & 3.52 & 0.8 & 
-22.05 & 3 \\
611 &  03:32:50.84 & -27:31:41.16 & SFG & --- & --- & --- & 
0.54 & 1.22 & 1.36 & --- & 
--- & 1 \\
612 &  03:32:50.86 & -28:03:17.64 & SFG & 0.10 & 21.34 & $<$40.57 & 
1.08 & 0.72 & 5.23 & --- & 
-18.91 & 3 \\
613 & --- & --- & SFG & --- & --- & --- & 
$<$-0.02 & --- & --- & --- & 
--- & 1 \\
614 &  03:32:51.59 & -27:59:18.46 & SFG & 0.91 & 23.15 & $<$42.42 & 
1.16 & 0.79 & 0.98 & --- & 
-22.85 & 1 \\
615 &  03:32:51.65 & -27:39:36.79 & RL AGN & 0.78 & 23.56 & --- & 
0.20 & 1.07 & 0.70 & 1.8 & 
-20.70 & 3 \\
616 &  03:32:51.73 & -27:49:51.02 & SFG & 0.74 & 22.89 & $<$41.80 & 
0.88 & 0.84 & 0.87 & --- & 
-22.00 & 3 \\
617 &  03:32:51.79 & -27:59:56.18 & SFG & 0.53 & 22.95 & $<$41.96 & 
0.78 & 0.90 & 1.15 & --- & 
-20.43 & 3 \\
618 &  03:32:51.84 & -27:44:36.78 & RQ AGN & 0.52 & 22.85 & $<$41.40 & 
1.08 & 1.01 & 2.02 & 0.1 & 
-22.29 & 2 \\
619 &  03:32:51.83 & -27:42:29.49 & RQ AGN & 1.03 & 23.44 & 42.58 & 
0.88 & 0.68 & 1.00 & --- & 
-23.18 & 3 \\
620 &  03:32:52.07 & -27:44:25.12 & RL AGN & 0.53 & 23.00 & 41.20 & 
$<$-0.80 & 0.48 & 0.38 & -0.1 & 
-22.37 & 3 \\
\hline
\end{tabular}
\label{tab_class}
\end{table*}

In Table \ref{tab_class} we report the classification for each radio source together with the information used to identify it. The identification number (column 1) corresponds to that given in \citet{miller13} and \citet{bonzini12}. The position of the optical-IR counterpart is given in column 2 and 3 \citep{bonzini12}. The table also includes the source redshift (5), the logarithm of the radio luminosity in W Hz$^{-1}$ (6), the logarithm of the unabsorbed X-ray luminosity in erg s$^{-1}$ (for undetected sources a 3$\sigma$ upper limit is given) (7), $q_{24obs}$ (8), the IRAC-colours (9 and 10), the radio spectral index (assumed to be $=0.7$ when not available) (11), and the rest frame absolute B magnitude (12).
We also define a quality flag (QF) which ranges from 3, for secure classification, to 1, for a tentative classification (column 13).
Sources for which all the criteria agree (or are not in contradiction) have QF=3 (45\% of the sources). 
We assign a QF=2 to sources with $q_{24obs}$ value just above the RL AGNs threshold ($q_{24obs,M82}-2\sigma$ $<q_{24obs}< q_{24obs,M82} -1\sigma$). Sources without redshift but with clear signature of AGN activity (e.g., in the Donley wedge or with very low values of $q_{24obs}$ value) have also QF=2. A total of 18\% of the sources have QF=2. 
The remaining sources have classification with QF=1. Examples of this last category are sources without redshift and without clear signature of AGN activity or without an optical/IR counterpart. Also sources with upper limits on the X-ray luminosity $>10^{42}$ erg s$^{-1}$, not in the Donley wedge, and above the RL AGN threshold have QF=1. These sources are classified as SFGs but, even if we can exclude that they are RL AGNs, there might be a contamination from RQ AGNs. Deeper X-ray observations are needed to discriminate between the two classes of sources.

\begin{figure}
	\centering
	\includegraphics[width=\columnwidth]{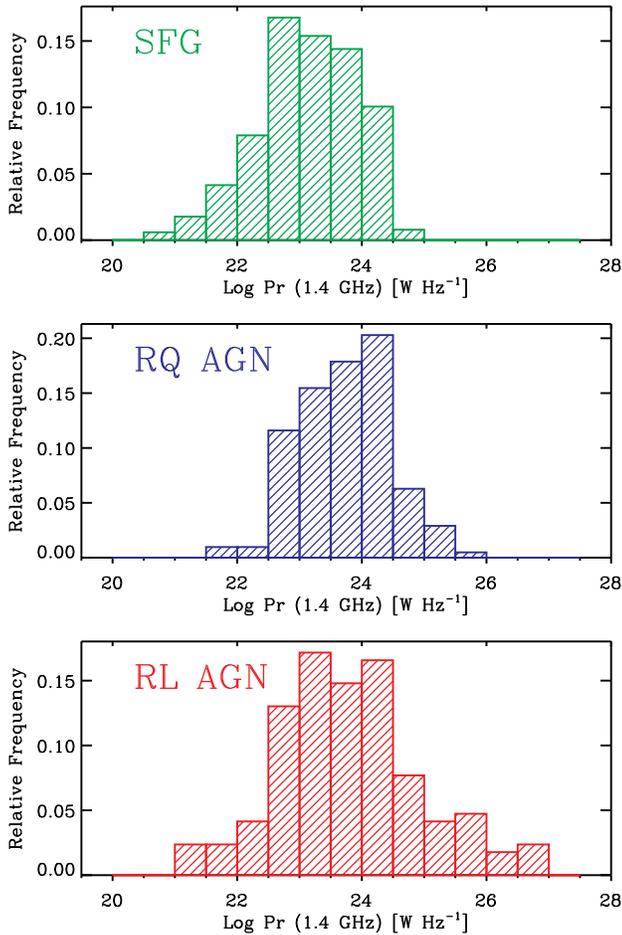}
 \caption{\small{Radio power distribution for SFGs (top), RQ AGN (middle), and RL AGN (bottom).}}
 \label{fig_Pr-pop}
\end{figure}

In Fig. \ref{fig_Pr-pop} we show the radio power distribution for the three classes of sources. The distribution of the RL AGNs is the widest since it includes both low power radio galaxies and the most powerful radio AGNs. 
This means that with a simple cut in $P_{\rm r}$ one would exclude a significant fraction of RL sources.
We also note that only 44 RL AGN out of a total of 177 have X-ray luminosity greater than $10^{42}$ erg s$^{-1}$ (35 sources) or are MID-IR selected AGN (9 sources). 
All the other are identified as AGN only based on their radio emission through the $q_{24obs}$ parameter.
We expect most of these X-ray undetected AGN to be intrinsically X-ray faint low power radio galaxies \citep[e.g.,][]{padovani11a}, but there could also be a fraction of heavily obscured sources \citep[e.g.,][]{delmoro13}. Radio observations are almost unaffected by dust extinction and therefore we can in principle detect even the most obscured systems.
The mean radio power for SFGs is $\sim 10^{23}$ W Hz$^{-1}$ but there are also galaxies with $P_{\rm r}$ as high as $10^{24.5}$ W Hz$^{-1}$. These sources have all redshift $>2$ but QF = 1.

We stress that our classification scheme tends to overestimate the number of SFGs. 
We consider AGNs only those sources that have strong evidences of the presence of an AGN in one of the wavelength range considered: radio, MIR, and X-ray. 

With this new scheme we confirm 80\% of the classification for the sample of 193 sources presented in  \citet{padovani11b}. The remaining sources were reclassified because of new redshift measurements, different optical/infrared counterparts, deeper X-ray data, and deeper 24 $\mu$m data, that allowed us to compute 
their $q$-value, whereas previously only an upper limit was available.

\section{The sub-mJy population}
\label{sec_number_counts}

\begin{figure}
\includegraphics[width=\columnwidth]{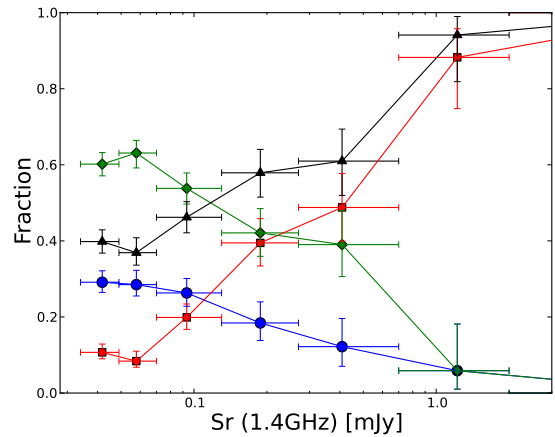}
\caption{Relative fraction of the
  various classes of radio sources: SFGs (green diamonds), all AGN (black triangles), radio-quiet AGN (blue circles), and radio-loud AGN
  (red squares).  Error bars correspond to $1\sigma$ Poisson errors \citep{gehrels86}.}
\label{fig_frac_classes}
\end{figure}

To investigate the relative contribution of the different source types to the radio population as a function of flux density, we exclude the outermost part of the field, since there we do not have enough ancillary multi-wavelength data to provide reasonable photometric redshift and therefore the classification of the sources is more uncertain.
We then consider the sub-sample within an area of 0.282 deg$^2$ where we have photometric redshift coverage \citep[see][for details]{bonzini12}, which includes 779 radio objects for which we have redshift for 90\% of the sources with optical/IR counterpart (compared to the 81\% of the whole radio sample).
We consider radio sources down to an rms noise of 6.5 $\mu$Jy, or a 5$\sigma$ flux density limit of 32.5 $\mu$Jy.  
The sensitivity of our sample is somewhat a function of the position in the field, although a much less strong one than for the CDFS sample of \citet{kellermann08}. Consequently, the area of the sky covered at any given flux density is flux density dependent 
\citep[see][]{miller13}. To estimate the relative fractions of sources as a function of flux density we therefore weigh each source by the inverse of the fraction of the area corresponding to that value. The results are plotted in Fig. \ref{fig_frac_classes}, which shows that AGNs dominate at large flux densities ($\ga 1$ mJy) but SFGs become the dominant population below $\approx 0.1$ mJy. 
Similarly, radio-loud AGNs are the predominant type of AGNs above 0.1 mJy but they drop fast at lower flux densities. 

In more detail, AGNs make up $43\pm4\%$ \citep[where the errors are based on binomial statistics:][]{gehrels86} of sub-millijansky sources and are seen to drop at lower flux densities, going from 100\% of the total at $\sim 10$ mJy down to $39\%$ at the survey limit. 
SFGs, on the other hand, which represent $57\pm3\%$ of the sub-millijansky sample, are missing at high flux densities but become the dominant population below $\approx 0.1$ mJy, reaching $61\%$ at the survey limit. 
Radio-quiet AGNs represent $26\pm6\%$ (or $60\%$ of all AGNs) of sub-millijansky sources but their fraction appears to increase at lower flux densities, where they make up $73\%$ of all AGN and $\approx 30\%$ of all sources at the survey limit, up from $\approx 6\%$ at $\approx 1$ mJy. 
These results are in good agreement whit those of \citet{padovani11b}. 

\section{Host galaxy properties}
\label{sec_host_gal}
\subsection{Morphology}
\label{sec_morphology}
A fraction of our sample has ACS/HST images available. We use this sub-sample to characterize the morphology of the host galaxy of our sources. 
We obtain the morphological information from the publicly available ACS-GC catalog \citep{griffith12}. 
The ACS images have been fitted with the GALAPAGOS method \citep{haeussler11} that analyses the images through the GALFIT code \citep{haeussler07}. 
The GALFIT code models each source with a S\`{e}rsic profile \citep{sersic63},
a parameter that describes the intensity profile of a galaxy, after a convolution with a point-spread function. In particular, we consider the results of the fit performed on the z-band (F850LP) ACS images. 
At the average redshift of our sample ($z=1.1$), this filter corresponds to rest-frame B band. 
We exclude from our analysis all the sources for which GALFIT gives an unreliable fit (FLAG=1 in the \citet{griffith12} catalog).
Moreover, sources with z-band magnitude $\geq 24$ suffer large uncertainties on the S\`{e}rsic index measurements and are therefore excluded. 
Finally we remove also low surface brightness galaxies since they are known to produce unreliable results \citep{griffith12}.

The total number of objects with good fit results is 362. Of these, 75 are RL AGNs, 60 RQ AGNs and 227 SFGs. The average redshift of this sub-sample is 0.76.
For these sources we have an estimate of the S\`{e}rsic index, which 
is generally lower for late type galaxies, where the disk dominates the intensity profile, and larger in elliptical galaxies.

\begin{figure}
	\centering
	\includegraphics[width=\columnwidth]{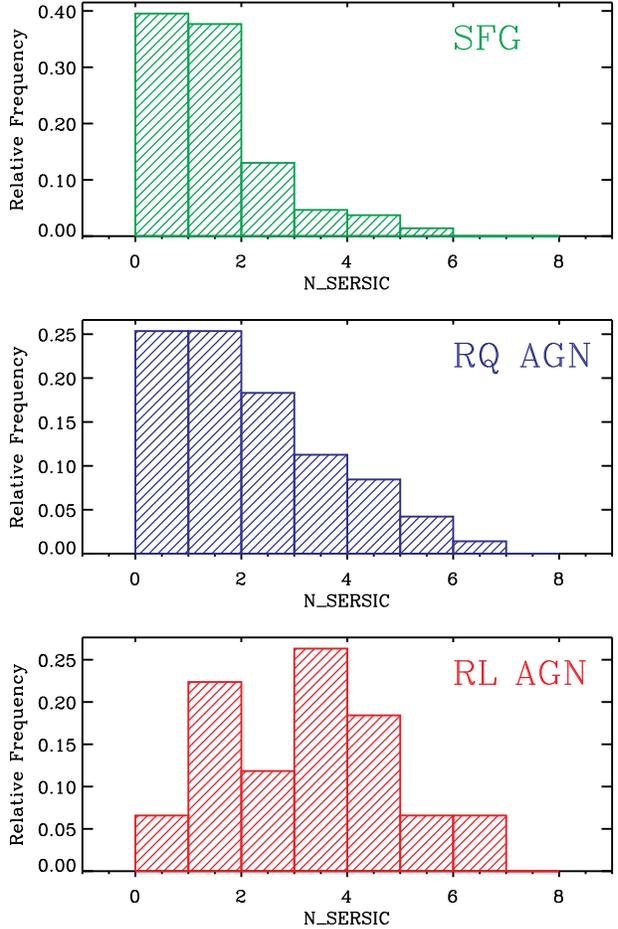}
 \caption{\small{S\`{e}rsic index distribution for SFGs (top), RQ AGN (middle), and RL AGN (bottom).}}
 \label{fig_sersic-histo-pop}
\end{figure}

The S\`{e}rsic index distribution for the three classes is shown in Fig. \ref{fig_sersic-histo-pop}.
The majority of RQ AGNs and SFGs have S\`{e}rsic indexes $< 2$ implying that the host galaxies of both classes are preferentially late-type objects. 
The distribution for the RQ AGNs is broader than the one of SFG, but it is important to note that the results of the GALFIT fit tend to overestimate the S\`{e}rsic index in the presence of bright central point source \citep{gabor09}. 
Therefore, the brightest QSOs among the RQ AGNs can produce a tail at high S\`{e}rsic indexes in the RQ distribution.
RL AGNs S\`{e}rsic indexes instead peak around 4, a value typical of galactic bulges and early-type galaxies. 
Performing a Kolmogorov-Smirnov test (KS-test) we find a significant difference ($>99.9\%$) between the S\`{e}rsic index distribution of RL AGNs and the other two classes. 
This result suggests that the host galaxies of RL AGNs are morphologically different from the host galaxies of RQ AGN and SFG: RL AGN are preferentially hosted in elliptical galaxies while RQ AGN and SFG are found in late-type galaxies. 

\subsection{Stellar masses}
\label{sec_masses}
To estimate the stellar mass of the host galaxies we used an SED fitting technique.
We modeled the observed photometry with two components: a galactic and an AGN component as in \citet{bongiorno12}.
For the AGN component we use the \citet{richards06} mean QSO SED. 
We applied an SMC-like dust-reddening law \citep{prevot84} and we consider different amount of dust extinction, $E_{AGN}(B-V)$, from $0$ to model  unobscured, type I AGN, to $9$ for the most obscured, type II sources, in steps of 0.1.
For the galactic component we use the stellar population synthesis models of \citet{bruzual03}. Assuming a universal initial mass function from \citet{chabrier03}, we generate SEDs assuming different star formation histories (SFH): ten exponentially declining SFH ($SFR\propto e^{-Age/\tau}$) with e-folding times ($\tau$) ranging from 0.1 to 30 Gyr, four models with constant SFR (1, 5, 10, 50 $M_{\odot}$ yr$^{-1}$) and five models with rising SFH ($\tau= -0.5, -1, -5, -10 , -15$). 
For each SFH, we generate SEDs for different ages from 50 Myr to 9 Gyr.
We exclude the models with age larger than the age of the Universe at the source redshift.
For the galactic component we assume a Calzetti's reddening law \citep{calzetti00}.
We consider colour excess $E_{gal}(B-V)$ in the range $0 \le E_{gal}(B-V) \le 0.5$ in steps of 0.1.
We impose the prior that $E_{gal}(B-V) <0.15$ if $Age/\tau>4$ to exclude models with large dust extinction in absence of a significant star formation rate \citep{fontana04,pozzetti07}.
The observed flux is modelled as the sum of the AGN and galactic component such as:
\begin{equation}
f_{obs}=af_{AGN} +bf_{GAL},
\end{equation}
where $a$ and $b$ are the normalization constants for the two templates.
The best fit template combination and normalization are found using a standard $\chi^2$ minimization.
The photometry for our radio sources is taken from the BVR selected \citet{cardamone10} catalogue. We matched their catalogue with the position of our counterparts using a searching radius of 0.2$\arcsec$ and we found 569 matches. For the remaining sources we use the photometry from the K selected \citet{taylor09} catalogue. We found 78 additional matches. 
Finally, for those sources not detected in both the BVR and K selected catalogues, we use the photometry from the \citet{damen11} IRAC selected SIMPLE catalogue. We found 145 more sources but only 24 of them have a redshift and can therefore be used in the fitting procedure.  

\begin{figure}
\includegraphics[width=\columnwidth]{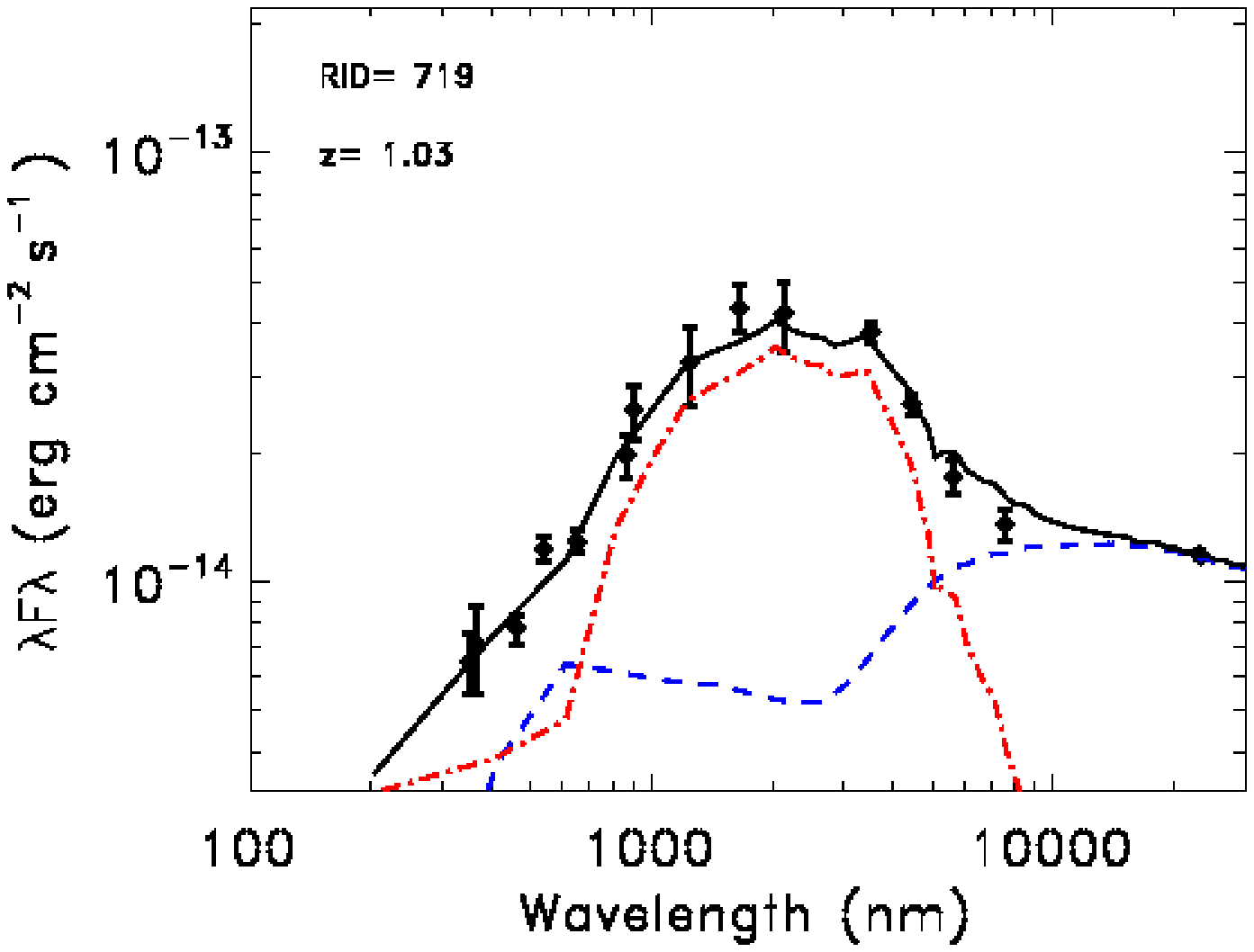}
\includegraphics[width=\columnwidth]{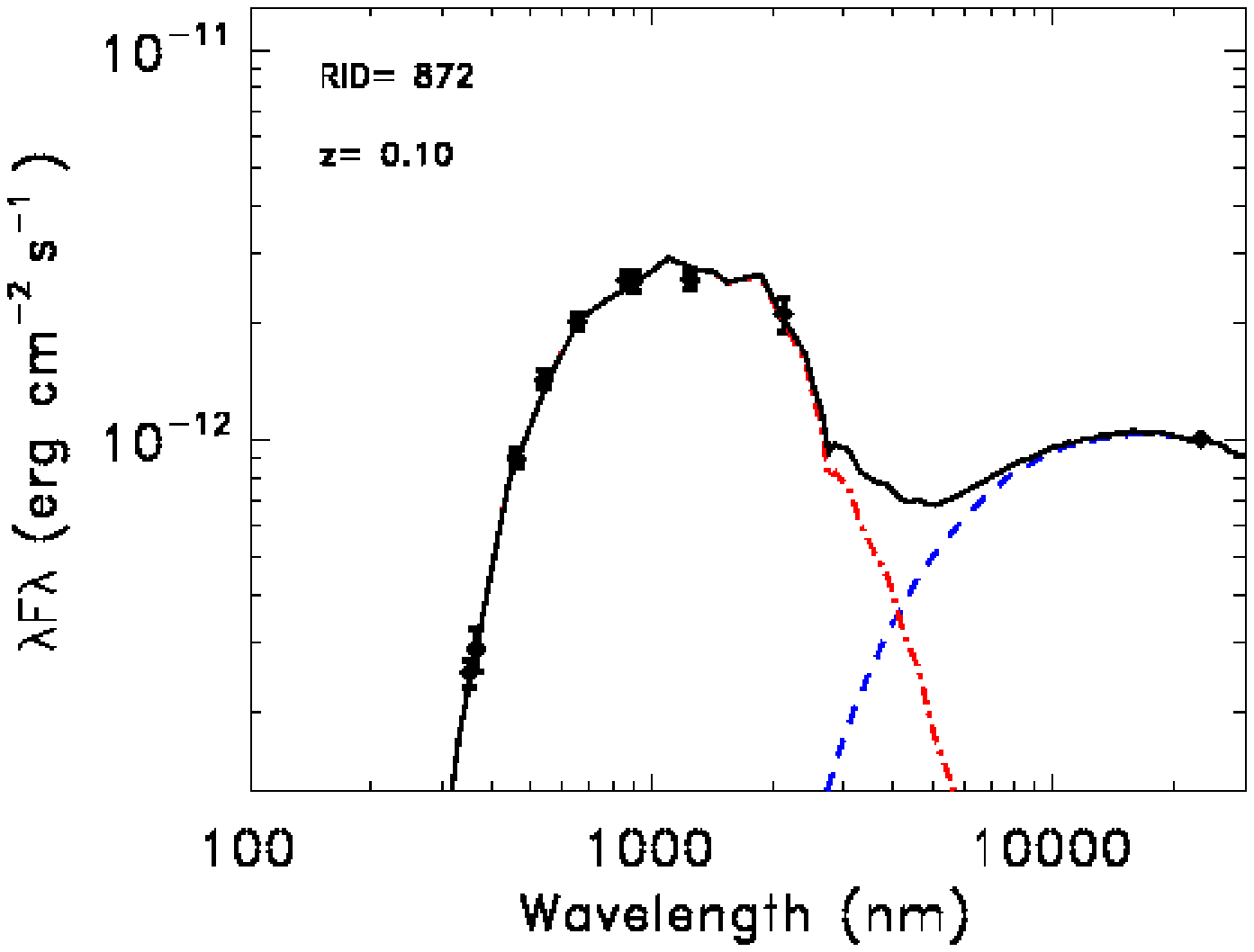}\\

 \caption{\small{Two examples of 2-component SED fit results. The galaxy component is plotted with the dot-dashed red line and the AGN component with the dashed blue line.}}
 \label{fig_best-fits}
\end{figure}

We fitted all sources with known redshift and at least eight photometric points in the wavelength range from the U-band to the 24$\mu$m for a total of 655 sources (23\% RL AGNs, 23\% RQ AGNs, 54\% SFGs). 
Two examples of the fitting results are shown in Fig. \ref{fig_best-fits}.

\begin{figure}
	\centering
	\includegraphics[width=\columnwidth]{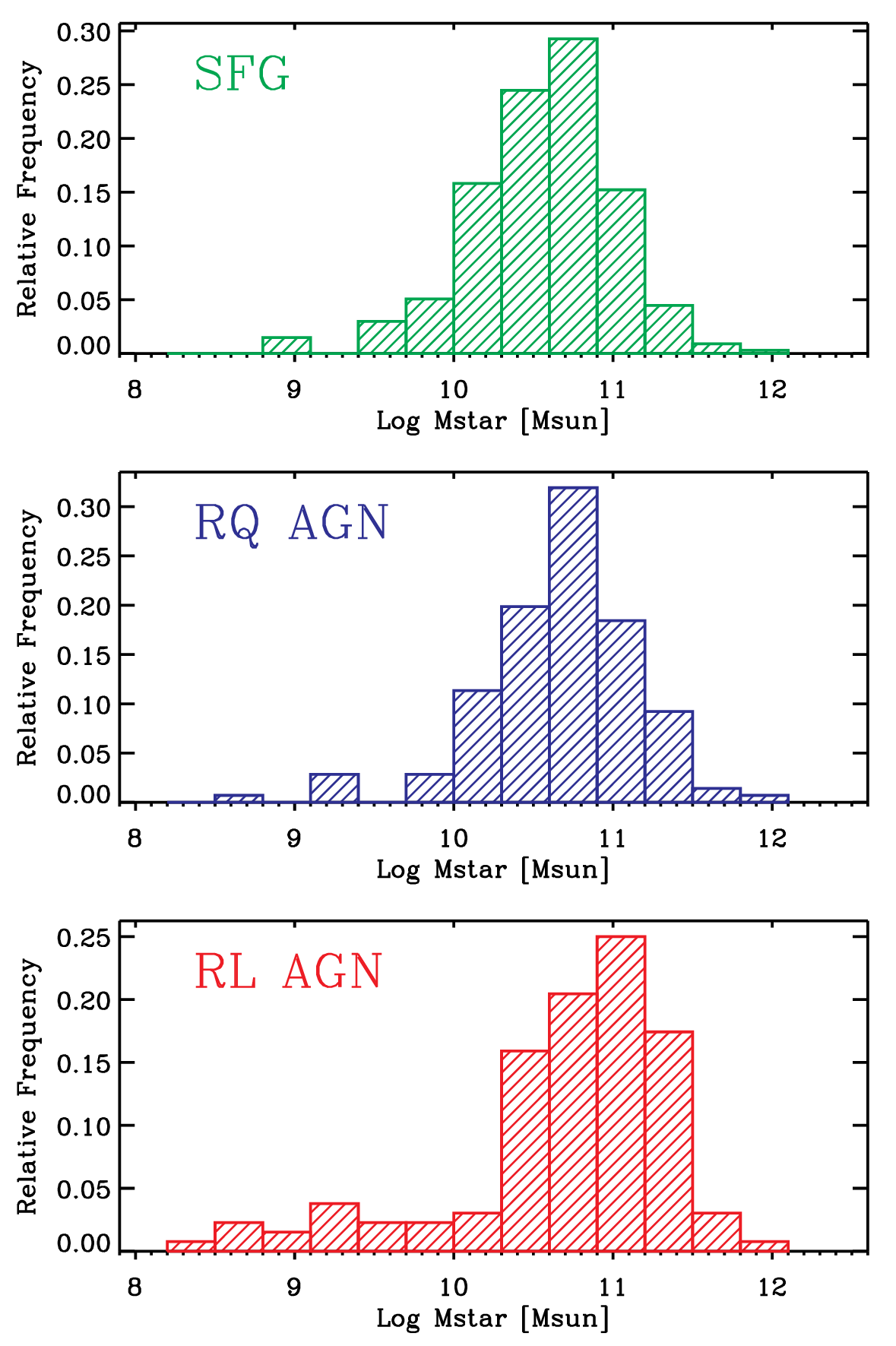}
 \caption{\small{Stellar mass distribution for SFGs (top), RQ AGN (middle), and RL AGN (bottom) in units of solar mass.}}
 \label{fig_Mstar-histo-pop}
\end{figure}

From the best fit galaxy model we derived the stellar mass of our galaxies.
We excluded 24 sources whose flux density in the rest-frame K-band is dominated by the AGN ($f_{AGN}>2f_{GAL}$), since the stellar mass measurement in these cases is too uncertain.
The stellar mass distributions for the three classes are plotted in Fig. \ref{fig_Mstar-histo-pop}.
Our radio selected SFGs have typical stellar masses of $10^{10.5} M_{\odot}$. 
The RL AGNs have on average higher stellar masses ($10^{11} M_{\odot}$) and a KS-test shows that the stellar mass distribution of RL AGNs is different from the SFGs one at the $>$99.9\% level.
RQ AGN host galaxies have stellar masses slightly higher than SFGs hosts. 
However, a KS-test shows that the mass distributions of SFGs and RQ AGNs are not significantly different. 
Moreover, it is important to note that most of our RQ AGNs (76\%) are identified as AGN as their total X-ray luminosity is above $10^{42}$ erg s$^{-1}$.
It has been suggested in recent works \citep[e.g.,][]{aird12,bongiorno12} that a threshold in X-ray luminosity introduces a bias towards higher host galaxy stellar masses. 
The probability of a galaxy hosting an AGN of a given Eddington ratio is independent of stellar mass and
the number density of AGNs increases for decreasing Eddington ratio.
Therefore in a flux density limited sample, one can detect AGNs of low Eddington ratios only in the most massive galaxies. 
Considering this bias, we cannot conclude that the host galaxy of RQ AGNs are intrinsically different from the SFGs ones. 

We also note that only 20\% of our RL AGNs have X-ray luminosity greater than $10^{42}$ erg s$^{-1}$ and therefore their high stellar masses cannot be explained by this effect. We then confirm the tendency of RL AGNs to be hosted in the most massive objects \citep[e.g.,][]{dunlop03,best05a}. 

\subsection{Rest Frame colours}
\label{sec_RF_colors}
\begin{figure*}
	\includegraphics[width=\columnwidth]{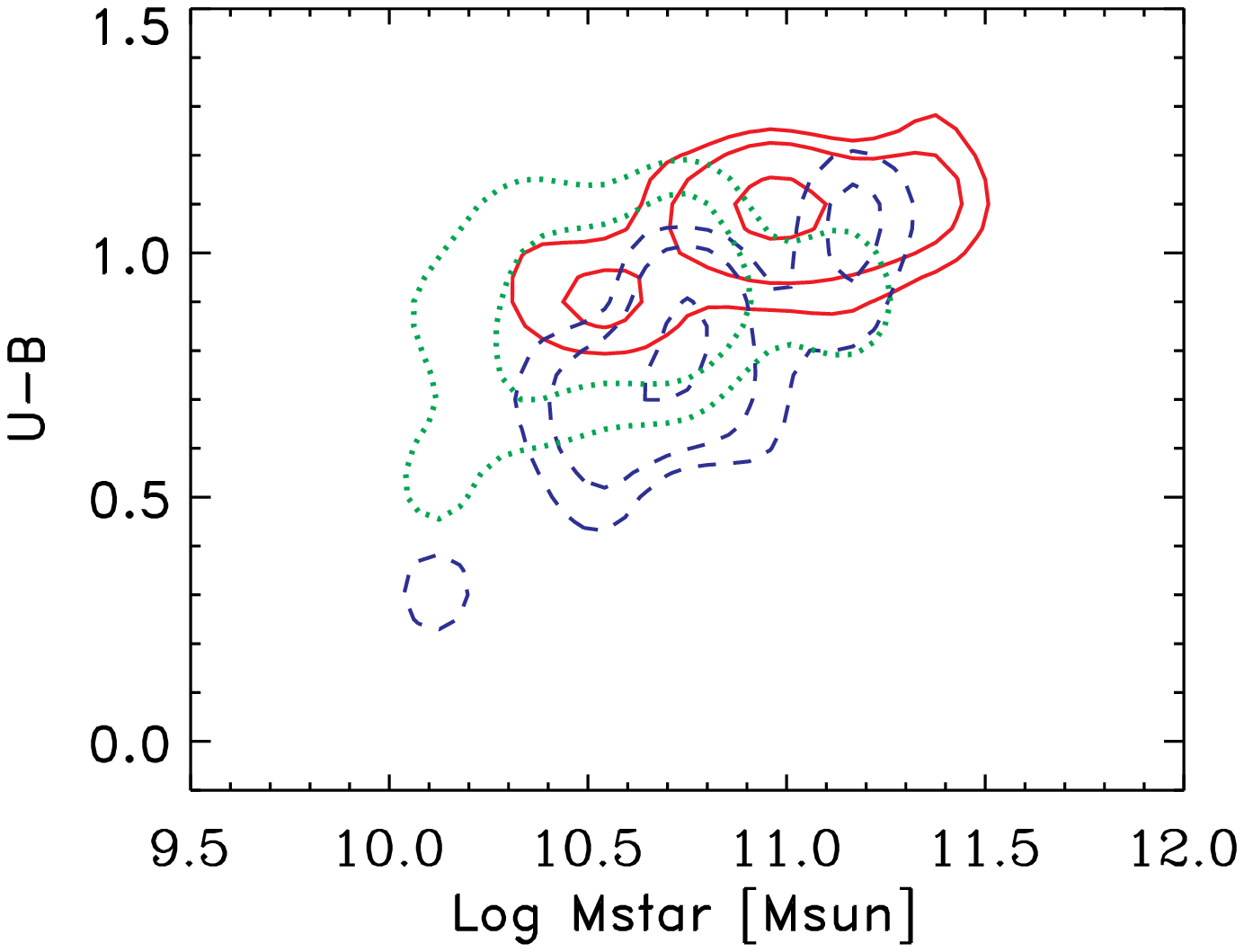}
	\includegraphics[width=\columnwidth]{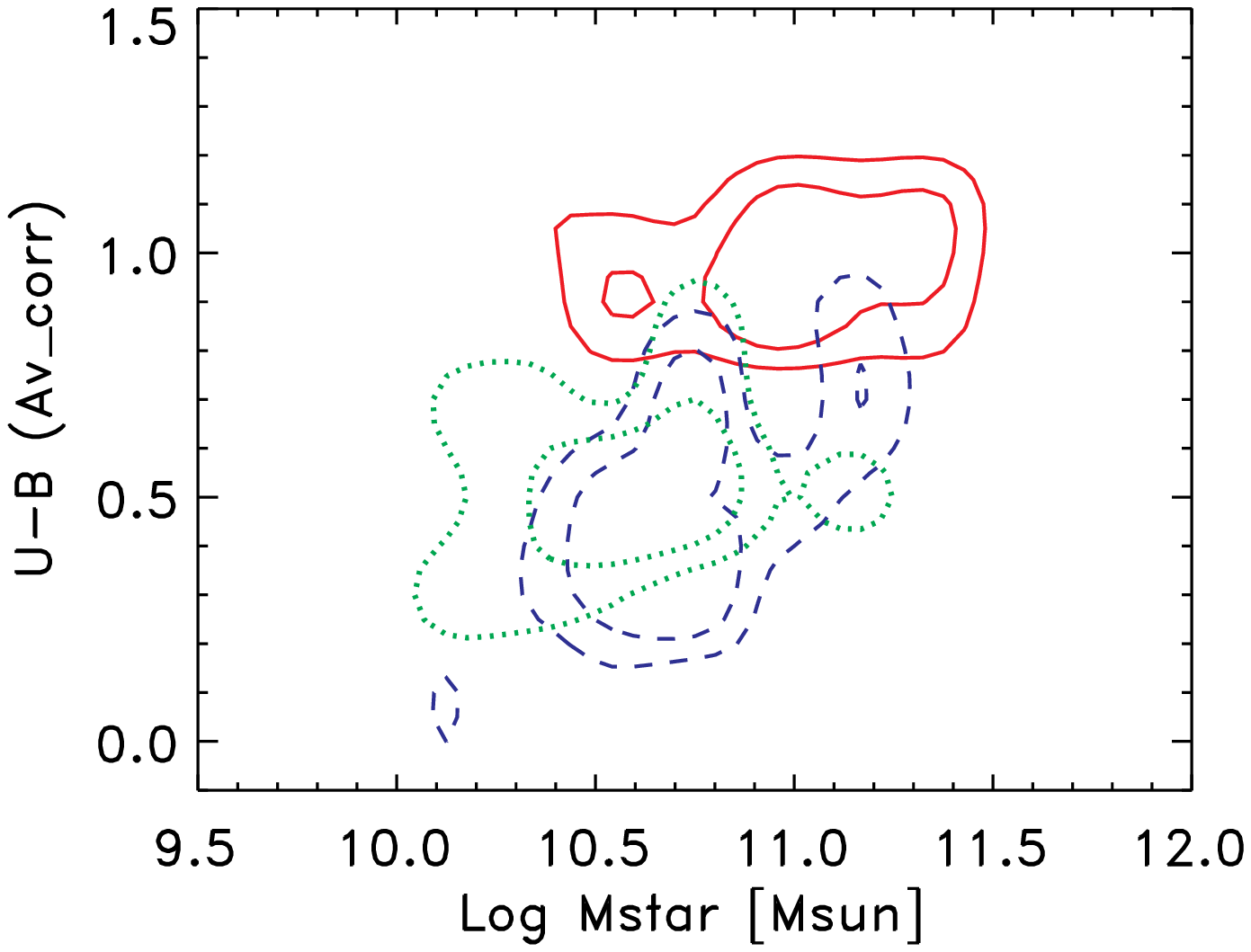}
 \caption{\small{Observed rest-frame U-B colours (left panel) and rest-frame U-B colours corrected for dust extinction (right panel) as a function of stellar mass for RL AGNs (red solid contours), RQ AGNs (blue dashed contours), and SFGs (green dotted contours).} }
 \label{fig_RF_col}
\end{figure*}

From the best-fit galaxy model we derived the rest frame U-B colours.
These colours are commonly used to distinguish between evolved stellar population galaxies, that populate the so called ``red sequence'' from young stellar population system usually found in the ``blue cloud'' \citep[e.g.,][]{bell04}.
In the left panel of Fig. \ref{fig_RF_col} we plot the measured rest frame U-B colours as a function of stellar mass. 
RL AGNs show preferentially red colours confirming previous results \citep[e.g.,][]{dunlop03}.
RQ AGNs and SFGs have instead a wider range of colours occupying  both the blue cloud and the red sequence, as well as the region in between, called the ``green valley.'' 
But the red colours in these type of systems should probably be interpreted as a signature of dust obscuration rather then as an indication of old stellar population. 
This can be tested by considering the de-reddened colours. The intrinsic colours are derived from the best-fit galaxy model corrected for dust extinction.
The latter are shown in the right panel of Fig. \ref{fig_RF_col} as a function of the stellar mass. 
Comparing the left and right panel, we see more clearly the impact of dust reddening: RL AGNs hosts, intrinsically characterized by old stellar population, are nicely displayed along the red sequence, while SFGs and RQ AGNs have moved towards bluer colours. 
In particular, it is interesting to note how the two type of radio AGNs have very different host galaxies properties, in agreement with what was found from considering their morphology (see Section \ref{sec_morphology}). 

\subsubsection{Rest-frame optical colours as classification method}
\citet{smolcic08} present a method to separate SFGs from AGNs, based on the rest-frame optical colours. This makes use of the rest frame colour P1 that is a linear combination of the colours obtained from the modified Str\"{o}mberg filters, as described in \citet{smolcic06}. In particular, according to \citet{smolcic08}, a source with P1 $<$ 0.15 is classified as SFG while it is an AGN otherwise. 
This method does not apply to QSO defined as point-like sources in the optical image \citep{smolcic08}. Therefore, we concentrate on the sub-sample described in Sec.\ref{sec_morphology} for which the results of the GALFIT analysis \citet{griffith12} are available and with known redshift. After removing the point-like sources we are left with 336  objects.   
We computed the rest-frame P1 colour for these sources from the best fit of the SED as in \citet{smolcic08}. We find that this method recovers about $70$\% of our RL AGNs, but only 25\% of our RQ AGNs. 
As described in Section \ref{scheme}, our RQ AGNs show clear evidences of AGN activity in the X-ray or in the MIR. The reason why these sources can easily be misclassified by a method based on optical colours is that the host galaxy can dominate the total emission at these wavelength.
RQ and RL AGNs have, on average, different host galaxy properties (see Sec. \ref{sec_host_gal}); therefore using the rest-frame optical colours it is possible to identify the RL AGN hosted in red, old galaxies, while it is hard to find RQ AGNs as they share the same parameter space as SFGs.

\subsection{The 4000 \AA\ break}
\begin{figure}
	\includegraphics[width=\columnwidth]{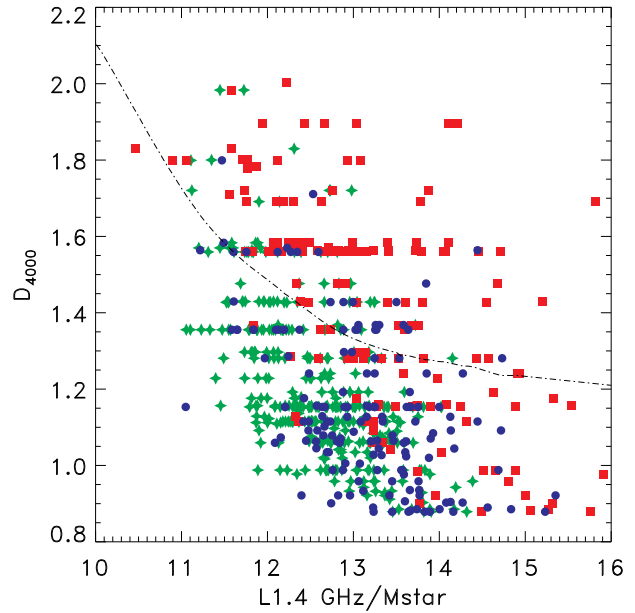}
 \caption{\small{Strength of the 4000 \AA\ break as a function of the 1.4 GHz luminosity normalized by stellar mass for RL AGNs (red squares), RQ AGNs (blue circles), and SFGs (green crosses). The dashed line correspond to the separation between RL AGN and SFG as defined in \citet{best05a}.} }
 \label{fig_Dn4000}
\end{figure}

The strength of the 4000 \AA\ break ($D_{4000}$) is a proxy for the mean stellar age of a galaxy. In \citet{best05a} it has been used to separate starburst galaxies from RL AGNs.
We computed the $D_{4000}$ for our sources from the best fit galaxy template obtained as described in Sec. \ref{sec_masses}. Deriving this quantity from the photometry, the uncertainties are large ($\sim 0.1$) as discussed in e.g., \citet{smolcic08}. In Fig. \ref{fig_Dn4000}, $D_{4000}$ is plotted as a function of the 1.4 GHz luminosity normalized by stellar mass. 
The dashed line marks the separation between RL AGNs and starburst galaxies as defined in \citet{best05a}.
The majority (65\%) of our RL AGNs lie above the separation line, in agreement with what expected for classical RL. The RL AGNs below this line are on average less powerful and the host masses are smaller. A total of 86\% of our SFGs are below the separation line. Given the large uncertainties on our $D_{4000}$ measurements due to the lack of spectroscopy, we can conclude that our classification method is in good agreement with the one proposed by \citet{best05a}. We confirm that the strength of the 4000 \AA\ break  can be successfully used to separate SFGs from RL AGNs, especially for sources with large stellar masses.  
Note that also in the $D_{4000}$ vs. $L_{1.4 GHz}/M*$, RQ AGNs share the same parameter space with SFGs. Indeed, 83\% of our RQ AGNs are below the starburst--AGN separation line, indicating the presence of a young stellar population in their hosts.
One more time we stress the need of a multi-wavelength approach to identify this population of AGNs, whose emission can be overshined by the host galaxy one or be heavily absorbed at optical wavelength.

\section{Discussion}
\label{sec_discussion}

\subsection{Selection caveats}
As already mentioned, the major source of uncertainty in our scheme comes from the possible misclassification of low luminosity RQ AGNs as SFGs where we do not have deep X-ray observations.
To get a rough estimate of the magnitude of this effect, we considered only the area covered by the 4 Ms Chandra observations (i.e., the 7$\arcmin$ radius around the center).
In this region we have a flux density limit for the hard X-ray band of $\sim 3.2 \times10^{-17}$erg cm$^{-2}$ s$^{-1}$ \citep{xue11} and we find 43 RQ AGNs and 63 SFGs.
Using the flux density limit of the 250 ks observations \citep[$\sim 5.5 \times10^{-17}$erg cm$^{-2}$ s$^{-1}$;][]{lehmer05} we would not detect 12 RQ AGNs that therefore would have been classified as SFGs having only an upper limit on their X-ray luminosity. 
From this comparison we can estimate that the contamination to the SFG population from RQ AGN in the outer part of the field is $\sim 16^{+7}_{-5}$\% (12/(63+12)).

A second source of contamination in our selection scheme can come from RL AGNs being classified as RQ AGNs due to strong 24$\mu$m emission related to AGN heated dust. 
Indeed a strong contribution to the 24$\mu$m from the AGN can boost the $q_{24obs}$ value in the SFGs/RQ AGNs locus at a given radio flux density. 
At $z\sim 1$, the mean redshift of our sample, the 24$\mu$m emission correspond to a rest frame 12$\mu$m emission.
To estimate the AGN contribution to the total flux density at this wavelength we use the correlation found by \citet{gandhi09} between the hard band (2-10 keV) X-ray luminosity and the MIR luminosity: $\log$  $L_{\rm MIR}$ =  $(-4.37\pm3.08)+(1.106\pm0.071)\log L_{\rm X}$, where $L_{\rm MIR}$ is the monochromatic luminosity at 12.3$\mu$m, in units of  $\lambda L_{\lambda}$ expressed in erg s$^{-1}$.
This relation holds for both obscured and unobscured AGNs \citep{gandhi09}.
We then subtract the corresponding flux density from the measured 24$\mu$m one and re-computed the $q_{24obs}$ value, corrected for the AGN emission, for all our RQ AGNs in the redshift range $0.85<z<1.15$ (20 objects). We find that only one object would be classified as RL AGN after the correction for the AGN emission of the $q_{24obs}$ value.
Therefore, we can estimate that the contamination of the RL AGN population from RQ AGN is $\approx$ 5\%.

\subsection{Host galaxies properties vs. radio loudness}
\label{sec_host-vs-Pr}
Focusing our attention on the AGN population, we can study the properties of AGN host galaxies as a function of radio loudness.

While we find that RL AGNs are preferentially hosted in high stellar mass galaxies, we do not observe a similar trend for RQ AGNs (Sect. \ref{sec_masses}).
Both the morphology and the rest-frame optical colours show a clear separation between the two types of AGNs. 
RL AGNs are preferentially found in elliptical galaxies (S\`{e}rsic index $\sim 4$), while RQ AGNs have usually lower S\`{e}rsic indexes typical of late-type galaxies (Section \ref{sec_morphology}).
In Section \ref{sec_RF_colors}, we computed the rest-frame optical colours from the results of the two component (AGN+galaxy) SED fitting. 
We note that since the rest-frame colours are computed from the best-fit galaxy template only, we minimize the contamination from the AGN emission itself \citep[see][for a more detailed discussion]{bongiorno12}. 
RL AGNs lie along the red sequence (see Fig. \ref{fig_RF_col}) suggesting that their host galaxies are old stellar population systems. 
If we look at the measured rest-frame U-B colours, we observe that RQ AGNs have a wider range of colours but the red objects are more likely dusty systems rather than old stellar population objects. 
This appears more clearly when we consider the dust extinction corrected colours (right panel of Fig. \ref{fig_RF_col}); the intrinsic rest-frame optical colours of RQ AGN are bluer compared to RL AGNs, implying the presence of young stars in the host galaxies of these AGNs. 
Also in the IRAC colour-colour diagram (Fig. \ref{fig_IRAC_colors_pop}), we see that RL AGNs are more concentrated in the bottom left part of the diagram.
Here, we expect to find passive systems, with a declining power law behaviour across the four IRAC bands (See elliptical colour-colour track in Fig. \ref{fig_IRAC_colors_pop}) .

Given these differences in the host galaxy properties, we suggest that the RQ and RL activity occurs at two different evolutionary stages of the BH-host galaxy co-evolution. RQ AGN are in an early phase, when the galaxy is gas rich, is still forming stars, has a young stellar population, and the AGN is efficiently accreting. The radio activity of the AGN occurs instead at later times when the galaxy is gas poor, the accretion on the BH is inefficient, the star formation in the host decreases, and the stars get older and redder.

RL AGN can be further divided in high and low excitation sources \citep[e.g.,][]{laing94,baldi08, hardcastle09}. The excitation level of a source is usually quantified using a combination of different line ratios \citep[e.g.,][]{best12}. Given the high redshift of our sample and the incomplete spectroscopic coverage, the excitation parameter can be computed only for very few ($<10$) RL AGNs.
For these reasons, a detailed discussion about the excitation level of our RL AGNs goes beyond the scope of this paper.
In \citet{best12}, the host galaxy properties of high and low excitation RL AGNs have been compared. Low excitation sources are found to have on average larger $D_{4000}$ and higher stellar masses compared to high excitation sources. 
Our sample of RL AGNs spans the whole range of $D_{4000}$ found in the \citet{best12} sample suggesting that it is a mixture of low and high excitation sources. However, given the large uncertainties of our measurement of the 4000 \AA break strength and the intrinsic large scatter in the $D_{4000}$ distribution for both low and high excitation sources, a quantitative estimate is not reliable.
More recently, it has been found that it is possible to separate low and high excitation radio-loud sources on the basis of their MIR luminosities \citep{gurkan13}. They propose an empirical 22 $\mu$m luminosity threshold of $5 \times 10^{43}$ erg s$^{-1}$ below which almost all the RL AGN are low excitation sources. We extrapolated the 22 $\mu$m luminosity from the sources photometry. We find that about 85\% of the RL AGN have MIR luminosity typical of low excitation sources.
However, this method has not yet been tested for a faint sample as the one presented in this paper. Therefore, this fraction can only considered a tentative estimate.  
Finally, we note the all the studies conducted so far on the excitation level of radio sources have been applied only to RL AGNs samples \citep[e.g.,][]{baldi08,hardcastle09,janssen12,gurkan13}. For a radio selected sample of RQ AGNs the distinction between low and high excitation sources would be difficult even having good quality optical spectra since the host galaxy can strongly contaminate the optical emission. Also the method based on the MIR luminosity can not be applied as assume no contribution from young stars at 22$\mu$m that is usually not true for the hosts of RQ AGNs.
Even if RQ AGNs host galaxies are more similar to the one of high excitation sources (e.g., blue colours, small masses), direct measurements of their excitation state have not yet been conducted.

\subsection{Radio emission in RQ AGN}
The origin of the radio emission in RQ AGNs has been debated for a long time.
It has been suggested that RQ AGN are a scaled version of RL AGN at lower radio power \citep[e.g.,][]{miller93} or that the major contribution to the radio emission in these system is due to the star formation in the host galaxy \citep[e.g.,][]{sopp91}.
In particular, in \citet{padovani11b} the latter hypothesis was supported by the study of the cosmological evolution and luminosity function. They found that both are significantly different for the two types of AGNs, while they are indistinguishable for SFGs and RQ AGNs.

In this paper, we had characterized the host galaxy properties for the three types of sources. We find that the host galaxy of RQ AGNs and our radio selected SFGs have many similarities.
Both RQ AGNs and SFGs have luminosity profiles preferentially with low S\`{e}rsic indexes ($n<2$) suggesting that, on average, both have disk-like morphology.
The rest frame optical U-B colours reveals that they are dusty, but intrinsically blue as seen from the comparison of the right and left panels of Fig. \ref{fig_RF_col}.
These results support the hypothesis that the origin of the radio emission is related to star formation processes in both SFGs and RQ AGNs, as proposed in \citet{padovani11b} and \citet{kimball11}.
This will be further discussed in coming papers, where we will investigate the star formation activity in the host galaxies of our radio selected RQ AGNs using Herschel data (M. Bonzini et al., in preparation).
We will also study the luminosity function and evolution for the different population of the larger sample considered in this work (P. Padovani et al., in preparation) to be compared with the results of \citet{padovani11b}.
 
\section{Summary and conclusions}
\label{sec_summary}
The sub-mJy radio population turns out to be a mixture of different kinds of sources due to the synchrotron emission from the relativistic particles accelerated either by the black hole, or associated with star formation processes. 

In this work we present a simple scheme, which expands upon that of \citet{padovani11b}, to disentangle the different populations of the sub-mJy radio sky: RL AGNs, RQ AGNs, and SFGs.
We have shown that the ratio between the IR and radio emission, parametrized by the $q_{24obs}$ value, is the key parameter needed to identify RL AGNs.
To differentiate between SFGs and RQ AGNs it is instead necessary to consider other AGN activity indicators, namely the X-ray luminosity and the MID-IR observed colours. 
Classification methods based only on optical properties are inefficient in finding RQ AGNs since their emission can be dominated by the host galaxy or be heavily absorbed at these wavelengths.
The simplicity of our scheme makes it suitable to be applied to large radio samples with ancillary X-ray and IR data, without the need of spectroscopic follow-up.

We confirm the increasing predominance at lower flux densities of SFGs, which make up $\sim 60\%$
of our sample down to $\sim 32~\mu$Jy and become the dominant radio population below $\approx 0.1$ mJy. RQ AGNs are also confirmed to be an important class of sub-millijansky sources, accounting for $ 26\%$ of the sample and $\sim 60\%$ of all AGNs, and outnumbering RL AGNs at $\la 0.1$ mJy.

We study the host galaxy properties of the three types of sources. Stellar masses and rest-frame optical colours of the galaxy are obtained using a two component SED fitting technique (galaxy+AGN) that allows us to subtract the AGN contribution.
Morphological properties are based on ACS/HST images.
We observe differences in the host galaxy properties of RL and RQ AGNs both in the rest frame optical colours and in the morphology. RL AGNs are preferentially hosted in more massive, red, elliptical galaxies, while RQ AGNs have typically stellar masses of $10^{10.5} M_{\odot}$, bluer colours and late type morphology.
This result is in agreement with what is found in previous works and support our classification method.
It also suggests that the radio activity associated with the black hole is linked to the properties of the host galaxy. One possible scenario is that they represent two different evolutionary stages of the BH-galaxy evolution.
In the RQ phase the radio emission from the AGN is low or even absent and the galaxy is young and still forming stars.
In a later stage, the radio activity of the AGN becomes more important as the galaxy gets older and stops forming stars.

Comparing our radio selected SFGs and the host galaxies of RQ AGNs we find many similarities. This result further supports the hypothesis that the radio emission in RQ AGNs predominately comes from the star formation processes in the host galaxy rather than from the BH activity, in agreement with \citet{padovani11b}, \citet{kimball11}. In an upcoming paper we will discuss further this topic (M. Bonzini et al., in preparation).
Based on this conclusion, we developed a powerful tool to disentangle the two radio emission mechanisms. Indeed, using the $q_{24obs}$ only, we can separate AGN powered radio emitters from systems where radio emission comes mostly from star formation.
This result can be applied to large radio samples such as the ones that existing and planned radio facilities will provide \citep[e.g.,][]{norris13}.  

\section*{Acknowledgments}
This work is based on observations with the National Radio Astronomy
 Observatory which is a facility of the National Science Foundation
 operated under cooperative agreement by Associated Universities,
 Inc. 
MB acknowledges the support of the International Max Planck Research School (IMPRS).
SV and PT acknowledge support under the contract ASI/INAF I/009/10/0.

\bibliography{BonziniBiblio}{}
\bibliographystyle{mn2e}


\label{lastpage}

\bsp
\end{document}